\documentclass[aps,preprint,groupedaddress]{revtex4-1}

\usepackage{graphicx}
\usepackage{color}
\usepackage{comment}
\usepackage{subfig}
\begin{document}

%\preprint{}

\title{\normalsize{Who creates trends in online social media: The crowd or opinion leaders?}}


\author{Leihan Zhang$^1$, Jichang Zhao$^{2,1,\star}$ and Ke Xu$^1$}
%\thanks{}
\affiliation{$^1$State Key Lab of Software Development Environment, Beihang University\\
$^2$School of Economics and Management, Beihang University\\
$^\star$Corresponding author: jichang@buaa.edu.cn}


\date{\today}

\begin{abstract}
Trends in online social media always reflect the collective attention of a vast number of individuals across the network. For example, Internet slang words can be ubiquitous because of social memes and online contagions in an extremely short period. From Weibo, a Twitter-like service in China, we find that the adoption of popular Internet slang words experiences two peaks in its temporal evolution, in which the former is relatively much lower than the latter. This interesting phenomenon in fact provides a decent window to disclose essential factors that drive the massive diffusion underlying trends in online social media. Specifically, the in-depth comparison between diffusions represented by different peaks suggests that more attention from the crowd at early stage of the propagation produces large-scale coverage, while the dominant participation of opinion leaders at the early stage just leads to popularity of small scope. Our results quantificationally challenge the conventional hypothesis of influentials. And the implications of these novel findings for marketing practice and influence maximization in social networks are also discussed.

\end{abstract}

%\pacs{}
%\keywords{online social media, information diffusion, trends, opinion leaders}

\maketitle

\section{Introduction}
\label{sub:intro}

For the ongoing growth of online social media, people are inundated with increasing various kinds of information in daily life. With each one being a social sensor, a vast number of individuals across the globe continually share news, statuses and sentiments through their online social networks. For instance, Twitter-like services help users instantly sense real-world events and spontaneously voice their opinions. And the online social interactions like retweet, reply, comment and mention would then boost the information propagation, spread different ideas and synchronize the collective attention of massive individuals, which might finally produce trends in online social media~\cite{apnetwork,responsefunction}. That is to say, by replacing the traditional news portals and providing convenient channels for information exchange, online social media have fundamentally changed the diffusion patterns of social networks and brought about significant challenges to existing comprehensions. While at the same time, the big-data of behavioral records in online social media also provides an unprecedented chance to deeply investigate the detailed dynamics of diffusion in social networks from diverse perspectives.

In the previous study, much attention from multidisciplinary fields have been devoted to understand the mechanism underlying popularity trends. Particularly, inspecting the properties of collective attention and diffusion principles of novel items attract the main focus in recent decades. For example, hashtags in Twitter and its equivalents are frequently used to unravel the generation mechanism of social memes by comparing items that succeed or fail in making the social popularity~\cite{lehmannca,changhashtag,diffusionofli,diffusionofli2,cumulativeeffect}. Lehmann et al.~\cite{lehmannca} focus on tracks of hashtags in Twitter and identify discrete classes of hashtags by their popularity evolution over time. They also find that exogenous factors are more important rather than epidemic spreading in establishing hashtag popularity. Bao et al.~\cite{cumulativeeffect} investigate the cumulative effect in information diffusion of Weibo and disclose that additional exposures cannot improve the probability of retweets. Meanwhile, tracking hot topics or emergent events is also an effective way to disclose the dynamics of collective attention or collective response~\cite{collectiveresponse}, which essentially drives the formation of trends or spikes~\cite{asurtrends,baoprediction,Wunovelty,Gomezinfluence,lininferring,romerotopoics,Bauckhageattention,sasaharaattention,Ferrara2013,collectiveresponse}. Romero et al.~\cite{romerotopoics} study the mechanics of information diffusion by comparing the spread process across different topics on Twitter. Bauckhage et al.~\cite{Bauckhageattention} investigate adoption patterns of 175 social media services and Web businesses using the data of Google Trends and unravel that collective attention on almost all services experiences a phase of accelerated growth followed by saturation and prolonged decline. Two types of trending topics in Twitter are indemnified by Ferrara et al.~\cite{Ferrara2013} and those with global popularity come from the major air traffic hubs in US. Besides hashtags and topic tracking, the neologism is another window to reveal intrinsic factors in collective attention~\cite{Swarupnorm,Eisensteinmapping,Bentleyword}. Swarup et al.~\cite{Swarupnorm} focus on linguistic innovation phase of lexical dynamics in social networks and find that the chance of existence of a norm is inversely related to innovation probability. Eisenstein et al.~\cite{Eisensteinmapping} conclude that the diffusion of neologisms is restricted to geographically compact areas and tend to spread from city to city. While the above studies mainly concentrate on temporal or spatial dynamics of trends in online social media and the discussion about roles of different users in the formation of popularity is missing. Hence it can be the first motivation of our study.

User influence or social capital~\cite{socialcapital} in social networks can be a convincing proxy to understand the successful diffusion of innovation ideas, neologisms and new products. The conventional diffusion theory states that a minority of people, called influentials, are generally considered as one of the most critical factors that affect information cascades~\cite{diffusiontheory} and the theory is also named as influentials hypothesis~\cite{wattschallenge}. With the help of these influentials, a large-scale attention might be achieved at a extremely low cost~\cite{Katztwostep,tippingpoint,Flynnleaders,Dholakiaconsumer,Iyengarleader,icwsm10cha,wuimpact,Kempenodes,Trusovusers}. Hence many previous studies focus on how to target influentials in social networks. Cha et al.~\cite{icwsm10cha} investigate the dynamics of user influence on time and topics based on three measures of indegree, retweets, and mentions. They find that most influentials can have great influence over a variety of topics. Kempe et al.~\cite{Kempenodes} propose an influence propagation model to study the problem of targeting initial influential nodes. However, whether the traditional two-step flow theory~\cite{Katztwostep} is applicable in online social networks is still a riddle. Besides, the role of influentials might be over-emphasized. For instance, Romero et al.~\cite{Romeropassivity} propose an algorithm which can determine users' influence and passivity according to their information forwarding activities and they demonstrate that high popularity do not always imply high influence. Domingos et al.~\cite{Domingoscustomers} insist that the key factors of determining influence are respectively the relationship among ordinary users and the readiness of the social network to accept a novel item. More directly, Watts and Dodds~\cite{wattschallenge,wattsnetworks} challenge the influentials hypothesis and point out that social epidemics tend to be driven by a critical mass of easily influenced individuals. Harrigan et al.~\cite{harrigancommunity} also find that it is the community structure not hubs that can substantially increase social contagion in Twitter. While measurable evidence from empirical data to carefully validate these controversial statements against the influentials assumption is still missing, which is the second motivation of our study.

\begin{figure}
\centering
\includegraphics[width=16cm]{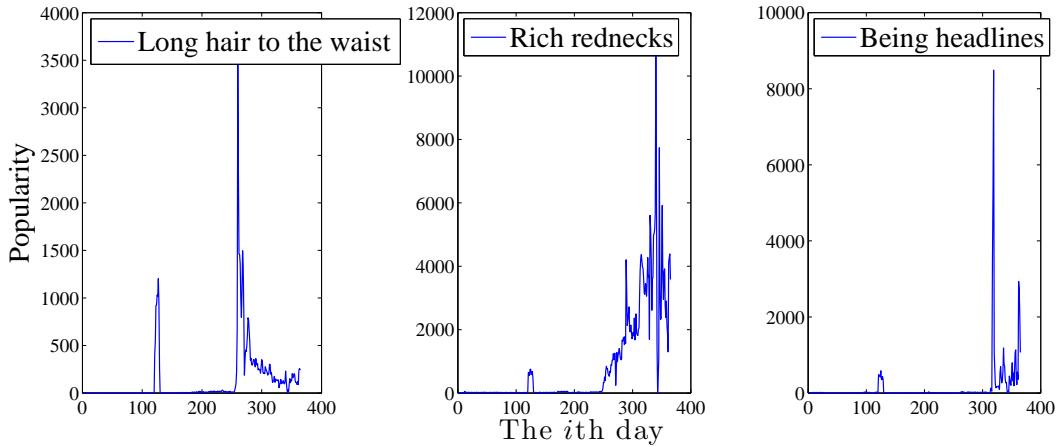}
\caption{Popularity trends of three randomly selected Internet slang words.}
\label{fig:2peakexample}
\end{figure}

Trends in online social media can be reflected by the popularity of hashtags, topics or even neologisms like Internet slang words. The collective attention underlying the popularity peaks indicate the participation of massive individuals during the diffusion of the corresponding novel items. As we have reviewed, the previous literature neglects the long-term analysis of popularity and mainly concentrates on the highest peaks. At the same time, study about the temporal dynamics of popularity is isolated from the discussion of roles that different individuals play in the diffusion. Aiming at filling these vital gaps, here we argue that disclosing how different individual functions in the propagation of an item should be embedded into the context of life-cycle dynamics of the popularity. Along this line, taking Weibo, the variant of Twitter in China as an example, we select typical 42 Internet slang words in the year of 2013 to observe how their popularities vary with time in this paper. And the popularity can be simply quantified by the occurrence frequency in daily tweets of Weibo. It is surprising that we find all these hot neologisms experience two peaks in their adoption history, in which the first one is much lower than the second one. Three examples are shown in Figure~\ref{fig:2peakexample} and this unexpected phenomenon does not draw attention in the previous study. In fact, two peaks indicate that in the life-cycle of each slang word, it possesses two opportunities to get the collective attention and then form a trend, and the first one fails but the second one succeeds. So the comparison of the diffusion represented by two peaks indeed offers us a decent window to understand how trends form in online social media. Specifically, the difference that differs the second peak from the first one is exactly the reason why it can obtain the collective attention and form a trend. Inspired by this finding, we try to figure out who actually create trends in online social media, the crowd constituted by ordinary users or opinion leaders with significant influence and great social capital.

\section{Results}
\label{sec:results}

Roles of individuals in diffusion can be divided into five segments according to their propensity of adopting a given innovation, including innovators, early adopters, early majorities, late majorities and laggards~\cite{diffusiontheory}. And among the five segments, early adopters are thought to be more critical than others~\cite{diffusiontheory}. So we can compare diffusions represented by two peaks of popularity in terms of the occupation of early adopters in each of them, since all tweets are randomly sampled from the stream.

For each slang word, we define its first popularity peak as $p1$ and the second peak as $p2$ and around these two peaks we can get two different diffusions, in which $p1$ represent the one with small-scale coverage while $p2$ represents the one with large-scale coverage. Assume the popularity value of a peak is $P$, we intercept the interval between $\alpha P$ and $\beta P$ ($0<|\alpha|\text{ and }|\beta|<1$) of the peak as $[\alpha,\beta]$, which is a specific stage of the diffusion that the peak stands for. As $\alpha<0$ and $\beta<0$, the interval is defined in the left half section of the peak, which is the early stage of the diffusion. Contrarily, the laggard stage of the diffusion can be defined in the right half section of the peak as $\alpha>0$ and $\beta>0$. While regarding to the user influence in online social networks, plenty of measures can be employed to reflect a user's significance to others in the network and among which, the number of followers (\#Followers) is generally a convincing indicator. Because the number of followers a user possesses directly determines how many potential listeners in the first step as the user posts a tweet in which a new word is adopted. Supposing each individual participates in adoption with the same probability, then more followers apparently indicate more spreaders in the second step. Cha et al.~\cite{icwsm10cha} also find that accounts with large numbers of followers are the most retweeted users in Twitter. Moreover, Wu et al.~\cite{wuimpact} evaluate user influence from five dimensions and show that the number of followers and their authority significantly affect the information in first-step communication. Hence here we use \#Followers to define the user influence and opinion leaders tend to possess great numbers of followers, while the crowd would only have a few. 

\begin{figure}
\center
\includegraphics[width=16cm]{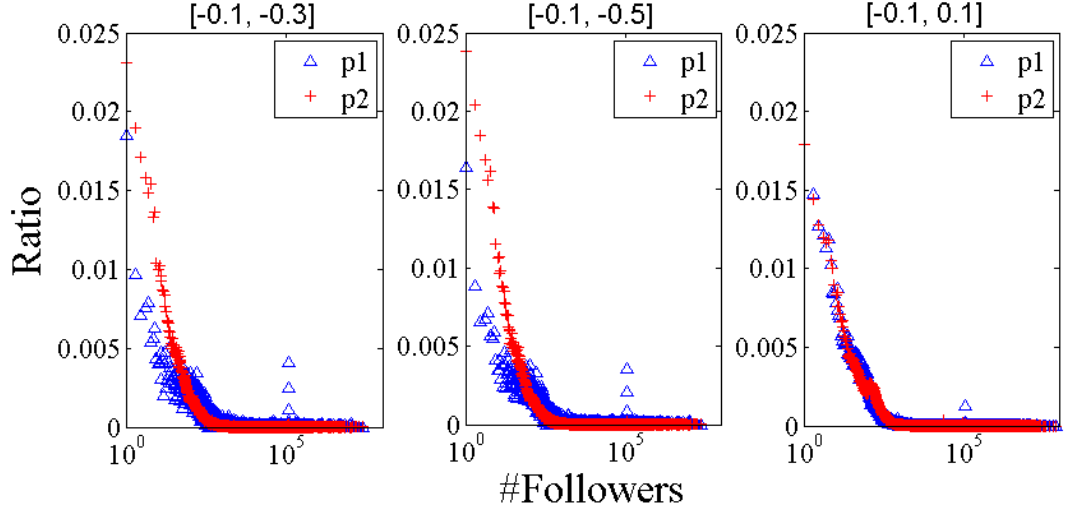}
\caption{The occupation of users with different numbers of followers in two diffusions. Three different diffusion stages are selected. All the results are averaged over the entire set of slang words we collect.}
\label{fig:followers}
\end{figure}

We first compare the occupation of different users with different numbers of followers at the early stage of two peaks and accordingly these users are the early adopters. As can be seen in Figure~\ref{fig:followers}, for the period of peaks ($[-0.1,0.1]$) there is no distinction between $p1$ and $p2$, indicating that the distribution of the number of followers for different participants at peaks is irrelevant to heights of the peaks. While in the early stage of diffusion ($[-0.1,-0.3]$ and $[-0.1,-0.5]$), the occupation of users with small \#Followers ($<200$), i.e. the crowd, is significantly higher than that of users with large \#Followers, i.e., the opinion leaders, in $p2$, which is the successful diffusion. Similarly, in all the stages of Figure~\ref{fig:followers}, we can observe a jump in the occupation of users with \#Followers around $10^5$ in $p1$, which represents the failed diffusion. The above observations are consistent and strongly suggest that the crowd occupy a higher proportion of the early adopters in $p2$ than that of $p1$, which indicates that the crowd's participation in the early stage of the propagation can produce a massive diffusion. On the contrary, domination of opinion leaders in the early stage cannot guarantee the success of the diffusion and its peak only reaches a small coverage like $p1$ shown in Figure~\ref{fig:2peakexample}.

\begin{figure}
\center
\includegraphics[width=16cm]{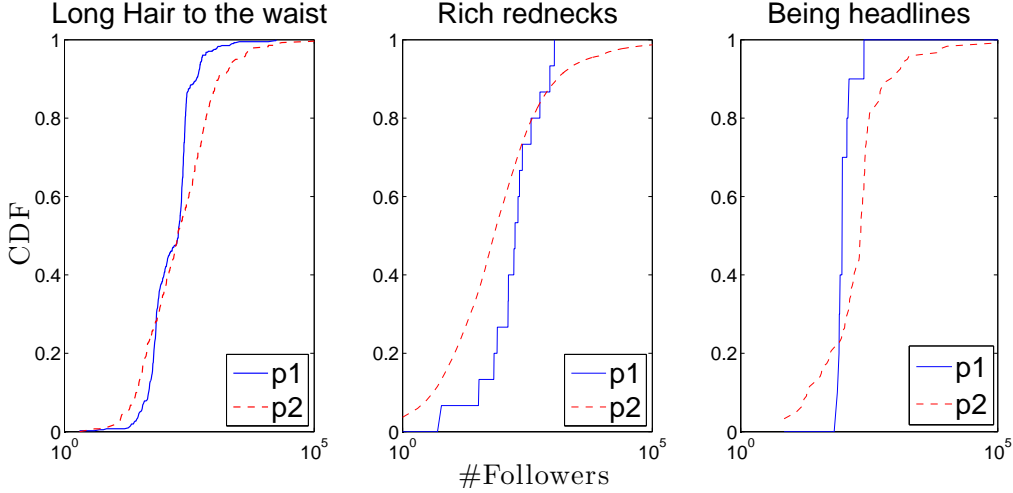}
\caption{CDF of \#Followers for three slang words.}
\label{fig:cdf}
\end{figure}

\begin{figure}
\center
\includegraphics[width=8cm]{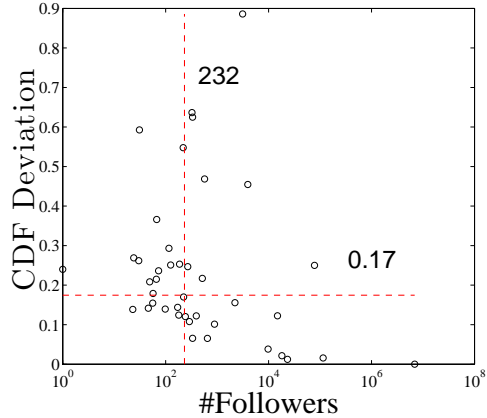}
\caption{The critical number of followers as CDF of \#Followers from different diffusions reaches the maximum deviation.}
\label{fig:threshold}
\end{figure}

From Figure~\ref{fig:followers} we can also find that during the early adoption segment, if the ratio of ordinary users with less than $\sim 200$ followers meets a critical condition, the diffusion would be more likely to obtain the collective attention of massive individuals. Hence we choose the early stage of $[-0.1,-0.5]$ to further explore the critical number of followers and the difference in occupations of different diffusions. In order the find the critical condition, we use the cumulative probability function (CDF) of \#Followers in early stages of different diffusions to determine the critical number of followers. As shown in Figure~\ref{fig:cdf} that the CDF curve of $p2$ exceeds the curve of $p1$ for small \#Followers. It again demonstrates the fact that ordinary users occupy greater proportion than opinion leaders for $p2$ and the point at which there is the greatest difference between these two CDF curves is the number we find. We define this number as a threshold and when users possess followers less than the threshold, their occupations of \#Followers in two peaks show the most significant deviation. The thresholds from 42 slang words are plotted in Figure~\ref{fig:threshold} and the median of the critical number of followers is 232 and the corresponding largest deviation of CDF is 0.17. It indicates that in the successful diffusion that $p2$ represents, the ordinary users with less than 232 followers occupy 0.17 higher prorogation than that in the failed diffusion represented by $p1$. Note that the threshold we find here is close to Dunbar's number~\cite{dunbarnumber}, which comes from the constrain of human cognition and still exists in online social networks~\cite{dunbar-zhao}. It is in agreement with the existing hypothesis that a global trend is beginning from many small-scale trends and the population number of an efficient group launching small-scale trends shall be less than Dunbar's number~\cite{tippingpoint}. And our findings also empirically testify the conjecture that the relationship among ordinary users is one of the key factors in determining influence~\cite{Domingoscustomers,wattschallenge}. Although the opinion leaders are important in starting the spread, but without enough participation of the crowd, the diffusion would probably fail in creating popular tends in online social media.

\begin{figure}
\center
\includegraphics[width=16cm]{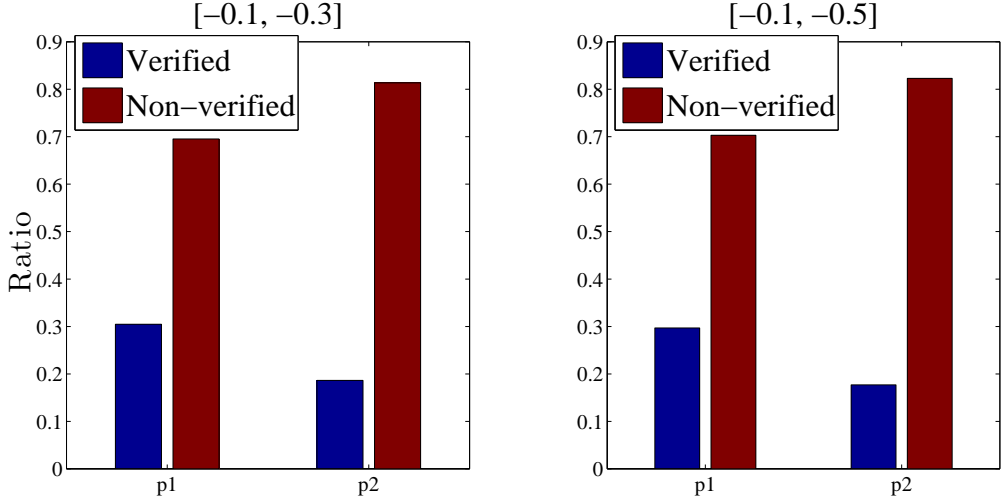}
\caption{Ratios of the verified and non-verified users in early stages of different diffusions.}
\label{fig:verified}
\end{figure}

The opinion leaders with great numbers of followers are always verified as VIPs by Weibo. As can be seen in Figure~\ref{fig:verified}, our further exploration shows that in both early stages of $[-0.1,-0.3]$ and $[-0.1,-0.5]$, the proportion of verified participants in diffusion represented by $p1$ is higher than that in the diffusion represented by $p2$. Specifically, the ratio between the verified user and the non-verified user is about $3:7$ in the small-scale diffusion with $p1$, while the ratio decreased to less than $2:8$ in the successful diffusion with $p2$. The fact suggests that in the process of a neologism becoming popular, non-verified users, which are almost ordinary users, play a decisive role in determining whether the phrase can attract the collective attention.

\begin{figure}
\centering
\includegraphics[width=16cm]{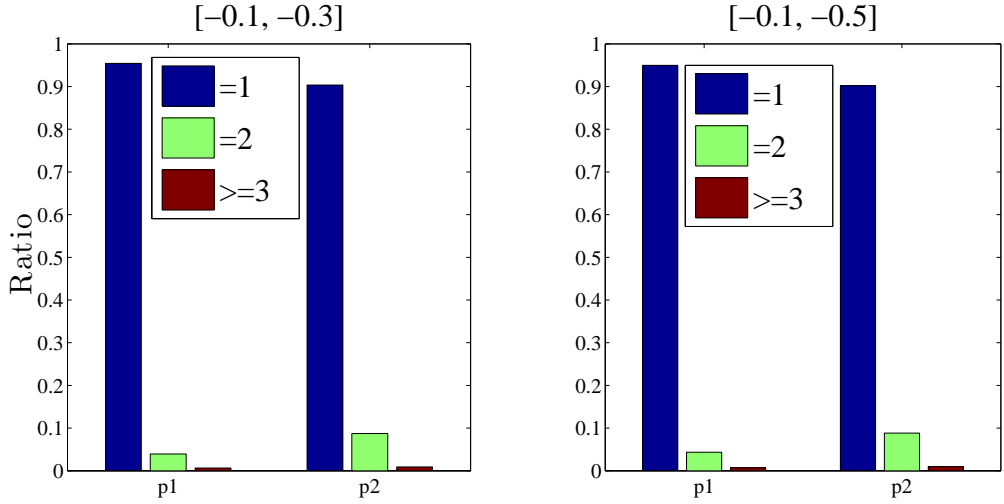}
\caption{Occupations of users with different retweet times in early stages of different diffusions. Here we mainly focus on times of 1 and 2, which take the majority of the retweets.}
\label{fig:retweet}
\end{figure}

Information diffuses in online social media mainly through retweet, which forwards the information from one user to its friends in the social network, especially in Twitter and its variants. As for opinion leaders in Weibo, because of larger numbers of followers, their tweets could possess much more first-step retweets than that of the crowd whose followers are less. However, only first-step retweets cannot propagate the information further in the network and second-time retweets play non-negligible roles in boosting the information spread. We compare the proportions of users with different retweet times in early stages of the diffusion and as can be seen in Figure~\ref{fig:retweet}, users with two-times retweets take higher occupation in the early stage of $p2$, which represents the successful diffusion. While the domination of one-time retweets in $p1$ does not make the diffusion reach high popularity. Though the forwarding of influentials might promote information to more followers at the first step, the repeat participation of the large amount of ordinary users can effectively facilitate the formation of trends.

\begin{figure}
\centering
\includegraphics[width=16cm]{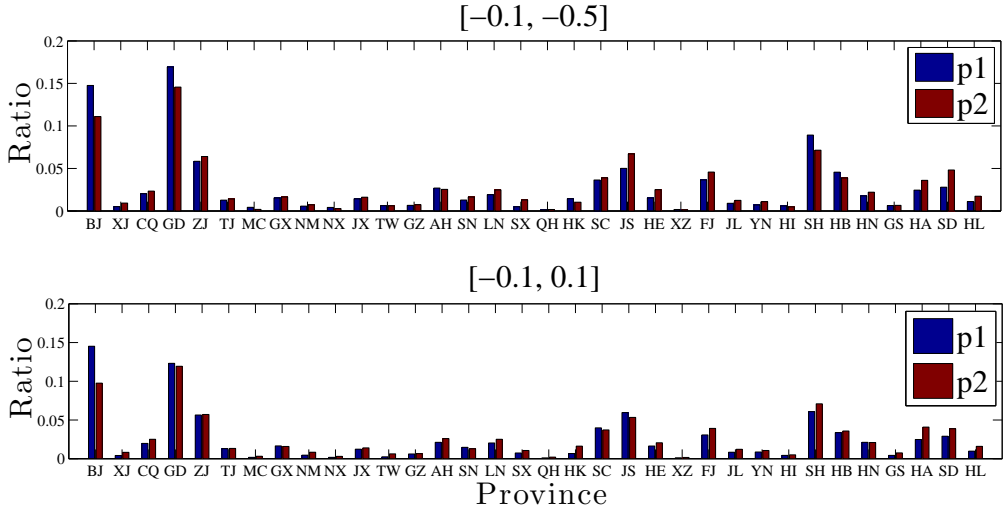}
\caption{Difference between geographical distributions of adopters at different stages of different diffusions.}
\label{fig:geo}
\end{figure}

Except the temporal dynamics, information also diffuses geographically in social networks, as found in [17] that neologisms tend to spread from city to city. So we also investigate the difference of geographical distribution of slang words' adopters in different stages of different diffusions. As shown in Figure~\ref{fig:geo}, for the early adoption stage of $[-0.1,-0.5]$, hot regions like Beijing (BJ), Guangdong (GD) and Shanghai(SH) occupy more users in the small-scale diffusion represented by $p1$ than that of the large-scale diffusion represented by $p2$. Specifically, the information entropy of the geographical distribution of early adopters is 4.26 for $p1$ but 4.46 for $p2$, which implies that early participants of the successful diffusion with $p2$ distribute more uniform across the country. It indicates that for the diffusion of $p1$, slang words are just local popular phrases and the opinion leaders from hot regions only produce the diffusion of limited coverage. We can also learn from Figure~\ref{fig:geo} that even for the period of peaks $[-0.1,0.1]$, the large-scale diffusion represented by $p2$ possesses greater information entropy (4.10) than that of $p1$ (3.72), indicating a more uniform and border adoption.

In summary, it is concluded that solid evidence from the in-depth comparison of two diffusions in the slang word adoption suggests an unexpected role of the crowd in creating trends. Different from the previous focus on influentials of social networks, we find the participation of ordinary users at the early stage can help the formation of trends in online social media.

\section{Discussion}
\label{sec:discussion}

Rapid growth of online social media in recent decades makes the online social network be the dominating channel for information exchange. For instance, Weibo in China has accumulated more than 500 million users in less than five years and it generates around 100 million tweets every day. These tweets deliver sophisticated signals of the real world, including personal statuses, public news, different opinions and diverse sentiments. Meanwhile, new items like events or neologisms might get the collective attention of massive individuals and form trends in extremely short period, which can be reflected through peaks in the popularity. 

Different from existing studies, in this paper we focus on long-term analysis of the popularity variation and find there are two peaks in the life-cycle of Internet slang words of 2013. Specifically, the first peak is lower than the second one and these two peaks can represent different diffusions of the same word, in which the former only reaches a small scope while the latter spreads globally across the network. Because of this, comparison of these two different diffusions could provide a decent window to investigate who creates trends, the crowd constituted by ordinary people or opinion leaders with great influence. Out of expectation, our results suggest that in online social media, the crowd play a decisive role at early stage of creating the trend, while the domination of opinion leaders in early adopters only produces a small-scale coverage. Contrary to the influentials hypothesis, opinion leaders can start a diffusion locally but only the participation of ordinary users can boost the spread to broad coverage and finally form a trend.

Our findings shed further insights into classical questions like viral marketing~\cite{leskovec:dynamics,Iyengarleader} and influence maximization~\cite{max-spread,chen-max-icdm2010} in social networks. The typical solution to these questions in the previous literature is mining the influentials to seed a diffusion. While results in this work suggest that more attention should be paid to the participation of ordinary users with followers less than 232 at the early stage. Given the definition of Dunbar's number, ordinary users in online social media usually stand for individuals capable of normal social interactions within the limit of human cognition ability. That is to say, online interactions between ordinary users and their behaviors are more reliable and diffusive than that of opinion leaders, who are always overwhelmed by too many social ties and act like inefficient hubs~\cite{harrigancommunity}. Because of this, focus on influential users only create a local and limited coverage, while attraction of the crowd could generate a massive diffusion with global coverage.

This study has important limitations. In comparison of different diffusions represented by different peaks, we mainly focus on the occupations of the crowd and opinion leaders. While the intrinsic factors that drive the crowd to participate in the adoption of slang words is not fully investigated. For example, it might be related to the slang word itself, pushed by the exogenous events or determined by the living environment (about one third of the first peaks happened on the May Day holiday, indicating that people are more likely to take part into the diffusion at leisure and comfortable time). In fact, Sasahara~\cite{sasaharaattention} even suggest that strong collective attention is accompanied by cascade of retweets and can be more related to the mood of users at the group level. Hence filtering out the essential factors underlying the participation of the crowd deserves further explorations in the future, which would help design appropriate strategy to attract adoption of the crowd at the early stage in reality. 

\section{Materials and Methods}
\label{sec:mm}

\emph{Data sets} The tweets employed in this study were collected through the open API of Weibo under its permit. We randomly sampled around 700 thousands tweets every day from the Weibo stream in the whole year of 2013. Each tweet contains attributes of ID, time stamp, text, retweet status and its author with the number of followers, address and verified state. We totally got 173,548,881 tweets and captured 9,021,435 users after spam filtering. The 42 popular Internet slang words in the year of 2013 were collected from the Chinese encyclopedia Website \url{baike.com}, which publishes a list of popular Internet slang words every year. The data set is publicly available to the research community and it can be downloaded freely through \url{http://goo.gl/WHXDjB}.

\emph{Identify peaks} We assume that the day with the highest popularity is where the second peak $p2$ locates and its popularity can be denoted as $P_2$. Then we search sequentially in the range between the date with popularity $0.01P_2$ (in the left side of $p2$) and the starting date and the date with highest popularity in this range is selected as the first peak $p1$.

\begin{acknowledgments}
This research was supported by 863 Program (Grant No. 2012AA011005), SKLSDE (Grant No. SKLSDE-2013ZX-06) and Research Fund for the Doctoral Program of Higher Education of China (Grant No. 20111102110019). Jichang Zhao was partially supported by the Fundamental Research Funds for the Central Universities (Grant No. YWF-14-JGXY-001).
\end{acknowledgments}

%\bibliographystyle{plain}
%\bibliography{refs}

\end{document}